%% file: main.tex
\documentclass[a4paper,twocolumn,11pt,accepted=2024-09-05]{quantumarticle}
\pdfoutput=1

\usepackage{makecell} 
\usepackage{graphicx}
\usepackage{dcolumn}
\usepackage{comment}
\usepackage{enotez}
\usepackage{braket}
\usepackage{caption}
\usepackage{subcaption}
\usepackage{ragged2e}
\DeclareCaptionJustification{justified}{\justifying}
\usepackage{bm}
\usepackage[utf8]{inputenc}
\usepackage[T1]{fontenc}
\usepackage{color}
\usepackage{tikz}
\usepackage{hyperref}
\usepackage{amsmath}
\usepackage{amssymb}
\usepackage{listings}
\usepackage{array}
\usepackage[ruled,vlined]{algorithm2e} 

\usetikzlibrary{shapes.geometric, arrows.meta, positioning, fit, calc}

\tikzset{
    startstop/.style={
        draw,
        rectangle, 
        rounded corners, 
        minimum width=3cm, 
        minimum height=1cm,
        align=center, 
        thick
    },
    process/.style={
        draw,
        rectangle, 
        minimum width=3cm, 
        minimum height=1cm, 
        align=center, 
        thick
    },
    decision/.style={
        draw,
        diamond, 
        minimum width=3cm, 
        minimum height=1cm, 
        align=center, 
        thick,
        aspect=2
    },
    arrow/.style={
        thick,
        -{Triangle[width=18pt,length=8pt]},
        rounded corners
    },
    note/.style={
        draw,
        align=left,
        font=\fontsize{10}{12}\selectfont
    }
}

\begin{document}


\title{SQuADDS: A validated design database and simulation workflow for superconducting qubit design}

\author{Sadman Ahmed Shanto}
\affiliation{Center for Quantum Information Science and Technology, University of Southern California}
\affiliation{Department of Physics  and Astronomy, University of Southern California}
\author{Andre Kuo}
\affiliation{Center for Quantum Information Science and Technology, University of Southern California}
\affiliation{Department of Physics  and Astronomy, University of Southern California}
\author{Clark Miyamoto}
\affiliation{Center for Quantum Information Science and Technology, University of Southern California}
\affiliation{Department of Physics and Astronomy, University of Southern California}
\author{Haimeng Zhang}
\affiliation{Center for Quantum Information Science and Technology, University of Southern California}
\affiliation{Department of Electrical and Computer Engineering, University of Southern California}%
\author{Vivek Maurya}
\affiliation{Center for Quantum Information Science and Technology, University of Southern California}
\affiliation{Department of Physics  and Astronomy, University of Southern California}
\author{Evangelos Vlachos}
\affiliation{Center for Quantum Information Science and Technology, University of Southern California}
\affiliation{Department of Physics  and Astronomy, University of Southern California}
\author{Malida Hecht}
\affiliation{Center for Quantum Information Science and Technology, University of Southern California}
\affiliation{Department of Physics  and Astronomy, University of Southern California}%
\author{Chung Wa Shum}
\affiliation{Center for Quantum Information Science and Technology, University of Southern California}
\affiliation{Department of Physics  and Astronomy, University of Southern California}%
\author{Eli M. Levenson-Falk}
\email{elevenso@usc.edu}
\affiliation{Center for Quantum Information Science and Technology, University of Southern California}
\affiliation{Department of Physics  and Astronomy, University of Southern California}%
\affiliation{Department of Electrical and Computer Engineering, University of Southern California}%
\maketitle

\begin{abstract}
We present an open-source database of superconducting quantum device designs that may be used as the starting point for customized devices. Each design can be generated programmatically using the open-source Qiskit Metal package, and simulated using finite-element electromagnetic solvers. We present a robust workflow for achieving high accuracy on design simulations. Many designs in the database are experimentally validated, showing excellent agreement between simulated and measured parameters. Our database includes a front-end interface that allows users to generate ``best-guess'' designs based on desired circuit parameters. This project lowers the barrier to entry for research groups seeking to make a new class of devices by providing them a well-characterized starting point from which to refine their designs.
\end{abstract}


Superconducting qubits are a leading quantum information technology platform. Scalable qubit fabrication requires accurate control of the Hamiltonian parameters most commonly used to predict device behavior, such as qubit anharmonicity and qubit-resonator coupling. This in turn requires accurate targeting of classical circuit parameters (inductances and capacitances). These are difficult to solve for, as there are typical no good analytical formulas (even approximate ones) to predict circuit parameters from design geometry. Instead, researchers must numerically solve the Maxwell equations given their design's unique boundary conditions.

Finite-element simulations of the electromagnetic field can provide reasonably accurate predictions of circuit parameters. However, these simulations are often time-consuming and require extensive computing resources for all but the most simple devices. They are also prone to misleading results, sometimes missing wildly on estimates of couplings between smaller features due to non-idealities such as finite mesh size in the solver. 
Even when simulations are perfectly accurate, there is still an essential problem: Hamiltonian parameters can be calculated from a device's design, but the inverse is not true. Design is thus a game of ``guess and check'' in order to find geometries that produce the desired Hamiltonian parameters.

Many resources have been developed in the community to assist with this design problem. Qiskit's Metal package allows for rapid programmatic generation of device layout \cite{noauthor_qiskit_nodate}. Packages such as scqubits, CircuitQ, and SQcircuit gives accurate numerical calculations of spectra, eigenstates, and/or Hamiltonians when given a lumped-element circuit model \cite{groszkowskiScqubitsPythonPackage2021, aumann2022circuitq, rajabzadehAnalysisArbitrarySuperconducting2023}. Techniques have been developed for modeling quantum circuits classically with lumped models such as SPICE \cite{tanamotoClassicalSPICESimulation2023}. Researchers have combined phenomenological rules with sophisticated optimization algorithms to better target Hamiltonian parameters \cite{yanEngineeringFrameworkOptimizing2020} and have even automated the search for optimal circuit element parameters \cite{menkeAutomatedDesignSuperconducting2021}. Some best-practices guides exist for device simulation, although they are typically not optimized for superconducting qubits \cite{mcdanielSimulationGuidelinesWideband2019}. However, the fundamental problem remains that there is no easy way to solve the problem that faces device designers: how does one generate a \emph{physical design layout} from the target effective Hamiltonian parameters.

In this paper we present SQuADDS: a \textbf{S}uperconducting \textbf{Qu}bit \textbf{A}nd \textbf{D}evice \textbf{D}esign and \textbf{S}imulation database. SQuADDS provides a robust workflow for generating and simulating superconducting quantum device designs. Our workflow is based on the Qiskit Metal library and is backed by an open-source database of well-simulated designs, including designs whose simulations have been validated by experimental measurements. A front-end interface for our database allows users to specify Hamiltonian parameters and rapidly generate ``best-guess'' designs, then perform accurate simulations of these designs to refine the design further if needed. SQuADDS is intended to function as a complement to the aforementioned circuit simulation codes, providing the physical layout to go along with the desired circuit parameters. Our database is designed as an open-source resource for the superconducting qubit community, with community members submitting experimentally-validated designs and simulation parameters. 

Typically, research groups will simulate designs, fabricate, test, and then iterate on the simulation parameters until the simulations converge to experimental results. The eventual working designs and simulation parameters are often held as ``secret sauce'' and not shared with the community. This presents a major barrier to entry for new groups, as fabrication and measurement runs are costly and time-consuming. Even for well-established groups it can be challenging to move to a new device type or geometry, as simulation parameters that worked for one style of device may not work for another. \emph{The purpose of the SQuADDS project is to remove this barrier, providing an open-source database pre-simulated designs and well-qualified simulation parameters.}

\section{Hamiltonian and Circuit Parameters}
In a typical experiment, a planar qubit is coupled to a transmission line resonator for readout or coupling to another element. Assuming a capacitive coupling and a charge type qubit, and neglecting offset charges, the Hamiltonian for such a circuit takes the familiar form
\begin{align}
   & H_r = 4 E_{C,r} n_r^2 + \frac{1}{2}E_L \phi_r^2 \\
   & H_q = 4 E_{C,q} n_q^2 - E_J \cos \phi_q \\
   & H_{int} = 4 e^2 \frac{C_c}{C_q C_r} n_q n_r \\
   & H = H_r + H_q + H_{int}
\end{align}
Here $E_{C,i} = e^2 / 2C_i$ is the charging energy on capacitor $C_i$, $n$ is the charge number operator, $E_L = \hbar/2 e L = \varphi_0^2/L$ is the inductive energy on resonator inductor $L$, $\phi_i$ is the phase operator, $E_J = \varphi_0 I_0$ is the Josephson energy, and $C_c$ is the coupling capacitance between qubit and resonator. All of these parameters are completely determined by the device geometry, with the exception of $E_J$ which is set both by geometry and fabrication parameters (i.e., tunnel barrier thickness).

It is often impractical to work with such circuit-level Hamiltonians. Following the standard quantization procedures and treating the qubit as a nonlinear oscillator (i.e., a transmon) yields the familiar Jaynes-Cummings model (with $\hbar = 1$) \cite{kochChargeinsensitiveQubitDesign2007,krantz2019quantum},
\begin{align} \label{eq:JCHam}
   H =  &\omega_r (a^\dag a + \frac{1}{2}) + \omega_q b^\dag b + \frac{\alpha}{2} b^\dag b(b^\dag b-1) \notag \\ 
   & + g (a - a^\dag) (b - b^\dag) 
\end{align}
Where 
\begin{align} \label{eq:g}
&\omega_r = \sqrt{8 E_L E_{C,r}} \notag \\
&\omega_q \approx \sqrt{8 E_J E_{C,q}} - E_{C,q} \notag \\
&\alpha \approx -E_{C,q} \notag \\
&g \approx \frac{C_c}{C_q}\sqrt{\frac{e^2\omega_r}{ C_r}}\left(\frac{E_J}{8E_{C,q}}\right)^{1/4}
\end{align}
Note that these expressions are approximations valid in the usual transmon limit $E_J>>E_{C,q}$ and the weak coupling limit $C_c<<C_q,C_r$. 

The qubit frequency $\omega_q$, anharmonicity $\alpha$, resonator frequency $\omega_r$, resonator linewidth $\kappa$, and qubit-resonator coupling strength $g$ are typical parameters of interest for superconducting qubit devices. Other parameters of interest, such as qubit-qubit coupling or Purcell filter linewidth, are discussed in Section \ref{sec:future}.

Beyond these approximations, many established resources exist in the community for numerically finding the exact Hamiltonian parameters from circuit parameters (capacitances and inductances). These includes the open-source codes scQubits, CircuitQ, and SQcircuit described above, based on rigorous Lagrangian analysis of the circuit \cite{kermanEfficientNumericalSimulation2020}. However, it is still a challenge to extract the circuit parameters themselves from the device design.

To extract circuit parameters, the standard technique is to use a finite-element solver such as COMSOL or ANSYS to simulate the electromagnetic field distributions. Performing accurate simulations requires careful tuning of the simulation parameters, including setting a fine enough discrete-element mesh and ensuring accurate convergence to a global minimum solution. Simulations are computationally intensive, requiring many hours or even days for larger designs even when run on high-performance computing clusters.

\section{Building the database}
\subsection{Validated designs}
\begin{figure*}[!ht]
    \centering
    \begin{minipage}[c]{0.4\textwidth}
        \includegraphics[width=\textwidth, height=2.5in, keepaspectratio, trim=0 1.5mm 0 0, clip]{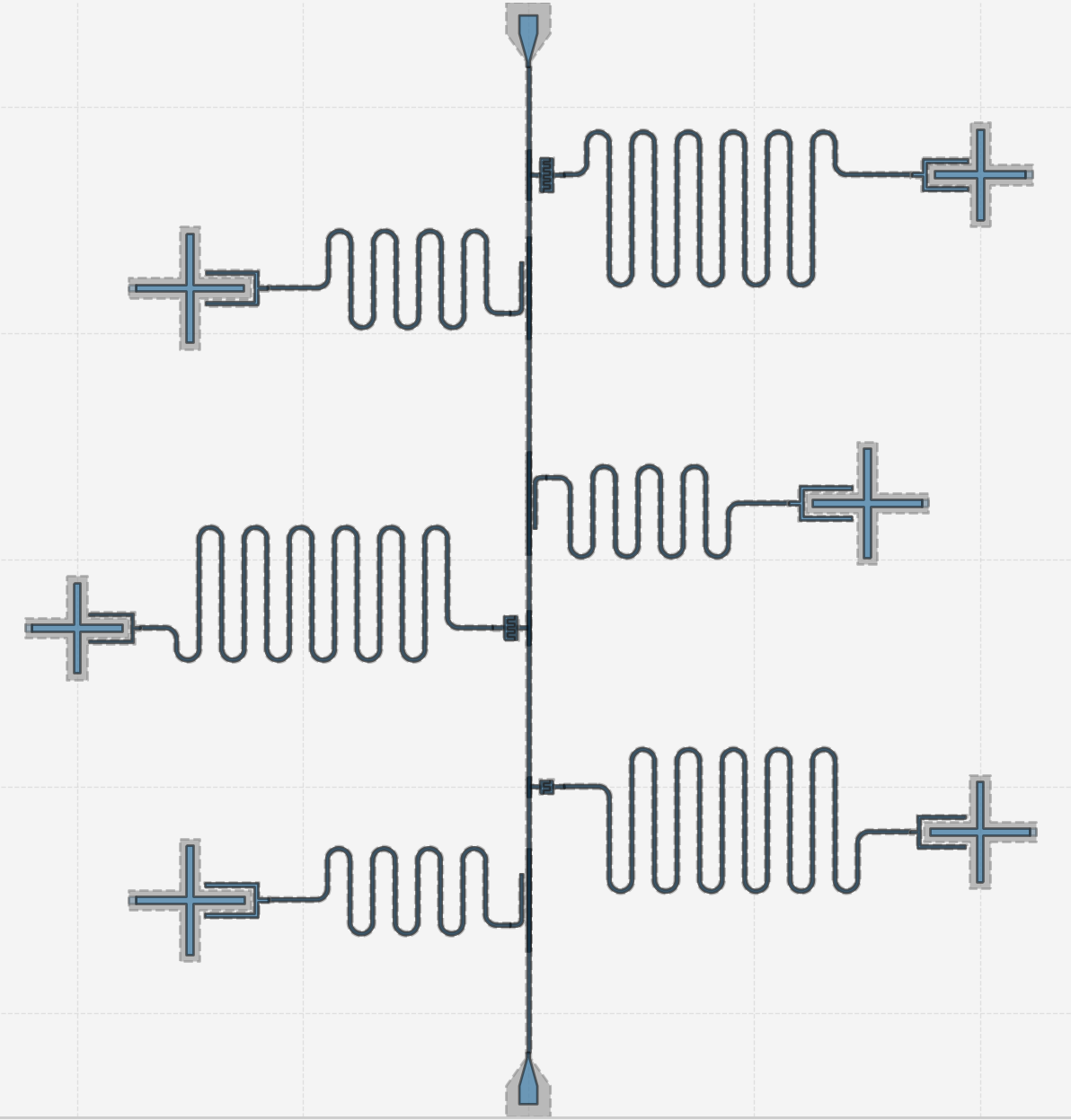}
        \label{fig:WM1design}
    \end{minipage}%
    \hfill
    \begin{minipage}[c]{0.6\textwidth}
        \centering
        \small 
        \begin{tabular}{cccccc}
            \hline
            \thead{\( f_{01} \) \\ (GHz)} & \thead{\( \alpha/2\pi \) \\ (MHz)} & \thead{\( f_{\text{res}} \) \\ (GHz)} & \thead{\( \kappa/2\pi \) \\ (MHz)} & \thead{\( \chi_{L} / 2\pi \) \\ (MHz)} & \thead{Extracted \( g/2\pi \) \\ (MHz)} \\
            \hline
            4.216 & -153 & 6.116 & 0.16672 & 1.56 & 60 \\
            3.896 & -154 & 6.353 & 0.18793 & 1.35 & 66 \\
            4.451 & -189 & 6.472 & 6.47625 & 1.97 & 70 \\
            3.586 & -164 & 6.568 & 0.21943 & 1.02 & 66 \\
            4.101 & -210 & 6.655 & 2.43003 & 0.82 & 52 \\
            3.881 & -176 & 6.704 & 0.78668 & 0.36 & 37 \\
            \hline
        \end{tabular}

    \end{minipage}
    \caption{ \label{fig:WM1}
    (a) Design layout of a chip with 6 xmon qubits, each connected to its own quarter- or half-wave readout resonator. The design is programmatically generated using Qiskit Metal. (b) Table showing the measured qubit frequency, qubit anharmonicity, resonator frequency, resonator linewidths, and Lamb shift, as well as the extracted \( g \) for a device with design similar to the one shown. All parameters are extracted from spectroscopic measurements.
    }
    \label{fig:WM1designPars}
\end{figure*}

The first step in building the SQuADDS database is compiling data from experimentally-measured devices. As an example, we show the design and experimental data for a chip with 6 independent xmon qubits, each coupled to its own half-wave or quarter-wave CPW resonator. See Fig.~\ref{fig:WM1designPars}. This device is a modification of the MIT Lincoln Laboratory ``standard candle'' chip and was fabricated via the MIT-LL SQUILL foundry program. To extract Hamiltonian parameters, we perform only two standard characterization measurements: ``punchout'' and qubit spectroscopy. In a punchout measurement, resonator spectroscopy is measured as a function of readout power; at high powers the circuit escapes from the Josephson energy well and is essentially erased from the system \cite{shillitoDynamicsTransmonIonization2022}. This punched-out resonance gives a measurement of the bare resonator frequency $\omega_r$ and linewidth $\kappa$, while the shift between the low-power and high-power resonances gives an accurate measurement of the Lamb shift $\chi_L \approx g^2/\Delta - g^2 / \Sigma$ where $\Delta \equiv \omega_r 
- \omega_q$ is the qubit-resonator detuning and $\Sigma \equiv \omega_r + \omega_q$ is the sum frequency. We note that the Lamb shift is typically derived in the rotating wave approximation (RWA), where the terms $ab$ and $a^\dag b^\dag$ in Eq.~\ref{eq:JCHam} are neglected; in this limit $\chi_L \approx g^2 / \Delta$, the formula typically reported in the literature. However, this approximation will significantly underestimate $g$ if $\Delta$ and $\Sigma$ are of the same order, which is common in weakly-coupled devices. This is the case for the example device, with qubit frequencies $\omega_q \sim 2\pi \times 3$ GHz and resonator frequencies $\omega_r \sim 2\pi\times 7$ GHz; taking the RWA with these parameters underestimates $g$ by approximately 25\%. In order to accurately translate between simulation and experiment, it is essential to include non-RWA terms in the Hamiltonian. A derivation of this Lamb shift and the dispersive shift given in Eq. \ref{eq:dispersive} is shown in Appendix \ref{app:shifts}.

A qubit spectroscopy measurement gives the qubit transition frequency $\omega_q = \omega_{01}$ and the frequency of the two-photon transition to the second excited state $\omega_{02}/2$. From this we extract qubit frequency anharmonicity $\alpha = \omega_{02} - 2\omega_{01}$. Combined with the punchout measurement, we can extract all Hamiltonian parameters. We note that these measurements are typically done at the very beginning of device tuneup for all experiments, and so this characterization adds no overhead.

In the case of devices where one mode is very lossy, spectroscopic measurements of that mode may be impossible. This is the case for the ``dissipator'' devices we include in the database, which have a tuneable coupler that is deliberately made lossy by coupling to an external feedline \cite{mauryaOndemandDrivenDissipation2024}. For these devices, we characterize the Hamiltonian parameters of the tunable-qubit Jaynes-Cummings model
\begin{align} \label{eq:tunableJC}
    H =  &\omega_r (a^\dag a +1/2) -\frac{\omega_q}{2}    (\cos^2\phi+d^2 \sin^2\phi)^{1/4}\notag \\
    &+ g(a+a^\dag)\sigma_x
\end{align}
Here $\phi = 2\pi \Phi_{ext} / \Phi_0$ is the external bias flux through the coupler SQUID in units of the reduced flux quantum, $d \equiv (E_{J2}-E_{J1})/(E_{J2}+E_{J1})$ characterizes any asymmetry between the SQUID junctions, and we take the approximation that the coupler is a qubit. In reality it is essentially a transmon, but the anharmonicity is difficult to measure in practice, and is much larger than $g$. However, as we have the ability to sweep the flux of the tunable coupler's SQUID and thus its $\omega_q$, we can directly measure the avoided crossing of the coupler and resonator levels with resonator spectroscopy. The resonator frequency as a function of flux is fit to 
\begin{align*}
&\omega_r'(\phi) = \\
&\frac{\sqrt{(\omega_r+\omega_q(\phi))^2+4g^2} \pm \sqrt{(\omega_r-\omega_q(\phi))^2+4g^2}}{2}
\end{align*}
to extract all Hamiltonian parameters in Eq.~\ref{eq:tunableJC} \cite{liu2023experimental}. 

\subsection{Matching Simulation to Experiment}
\begin{figure}[!htb]
\centering

\begin{subfigure}[t]{0.45\textwidth}
    \centering
    \includegraphics[width=\textwidth]{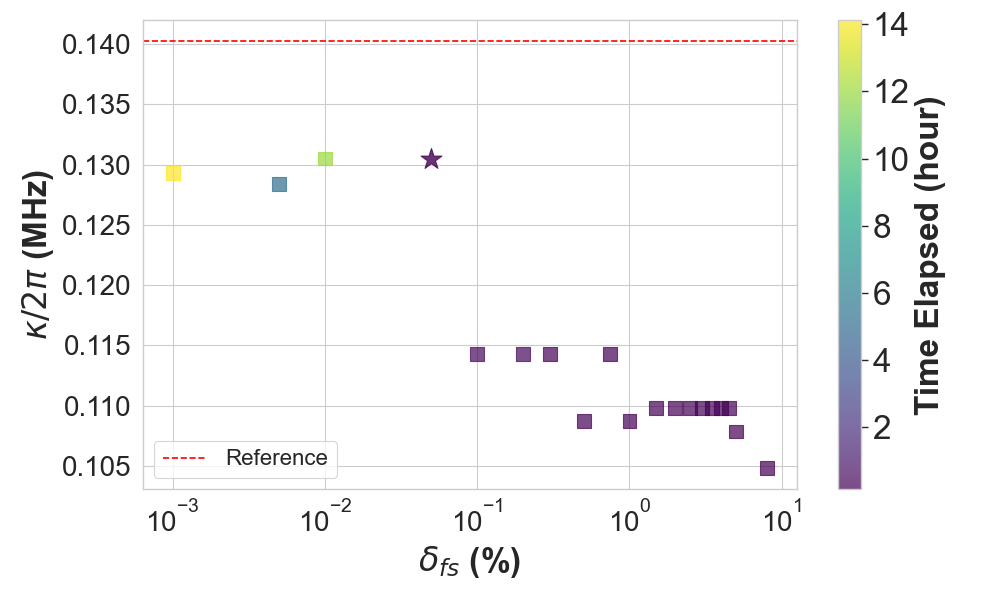}
    \caption{}
    \label{fig:panelA}
\end{subfigure}
\hfill
\begin{subfigure}[t]{0.45\textwidth}
    \centering
    \includegraphics[width=\textwidth]{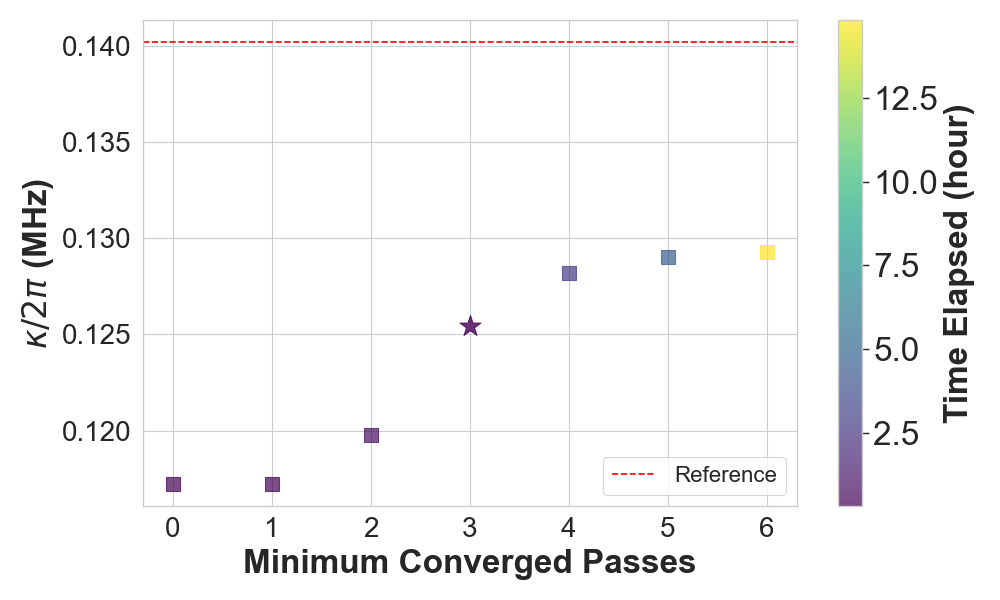}
    \caption{}
    \label{fig:panelB}
\end{subfigure}

\caption{\justifying
(a) Simulated resonator linewidth $\kappa$ as a function of convergence criteria, defined as the percentage change of the resonant frequency from one simulation pass to the next. Color indicates the time required for a simulation. The star symbol is the value $0.05\%$ that we use for simulations; further refinement does not significantly change $\kappa$ but does significantly increase simulation time. The dashed red line is the measured value of $\kappa$; a small systematic error persists, which is typical for this parameter and is likely due to impedance variations in the experimental device's environment. (b) $\kappa$ as a function of the minimum number of simulation passes below the convergence criteria. Again the star indicates the value 3 that we use for simulations. 
}
\label{fig:hyperparams}
\end{figure}

After collecting experimentally-measured Hamiltonian parameters for each design, we then generate the designs programmatically using Qiskit Metal. Occasionally this involves creating a new design element in Metal; we have contriubed several such elements to the Metal codebase. We then export the design to ANSYS for EM simulation using the HFSS solver. We use 2 solvers: the Q3D solver, which gives capacitance and inductance matrices for all elements, and the eigenmodal solver, which solves for the normal modes of the system and their quality factors. To ensure accurate results, we set simulation hyperparameters to mandate higher and higher accuracy (and thus take longer and longer times) until further refinement causes no significant change in the extracted Hamiltonian parameters. See Figure \ref{fig:hyperparams}. restrict the discrete mesh on each element to have a maximum size of $\lesssim 1/3$ the smallest characteristic dimension (e.g., the width of a strip of ground plane cross); in the case where a simulation uses the eigenmodal solver, we set the mesh on any junctions to be $\lesssim 1/10$ the characteristic dimension to ensure accurate modeling of the junction field. This is necessary for any future energy participation ratio (EPR) analysis of the circuit \cite{Minev2021EPR}, which we employ via the pyEPR package. We also set the convergence criteria to be a difference of less than $0.05\%$ from one simulation refinement iteration to the next, and we mandate at least 3 converged iterations. As shown in Figure \ref{fig:hyperparams}, even the most difficult-to-simulate parameter $\kappa$ converges to a saturating value that does not significantly change with longer, more precise simulations.

To extract parameters, we take the lumped-element capacitance matrices (if using Q3D) or the mode frequencies and quality factors (if using eigenmodal). If using lumped-element analysis, we plug the extracted capacitance and inductance values into the circuit-level Hamiltonian and use the scqubits code to extract qubit Hamiltonian parameters. Note that for distributed-element waveguide resonators this requires treating the resonator capacitance as an effective lumped capacitance
\begin{equation}\label{eq:rescap}
    C_r = \frac{\pi}{m\omega_r Z_c}
\end{equation}
where $\omega_r$ is the resonant frequency, $Z_c$ is the waveguide's characteristic impedance ($Z_c = 50$ $\Omega$ for all our designs), and $m=2$ for a half-wave resonator or $m=4$ for a quarter-wave resonator.

If using EPR analysis, we take the self-Kerr shifts and qubit frequency and extract the qubit charging energy $E_C$ using scqubits (method \texttt{Transmon.find\_EJ\_EC}), as for modest $E_J / E_C \lesssim 50$ the anharmonicity can diverge from $E_C$ by as much as 10\%. We likewise extract $g$ from the dispersive (cross-Kerr) shift that pyEPR reports using the formula derived from second-order perturbation theory:

\begin{equation}\label{eq:dispersive}
\chi = 2 g^2 \left(\frac{\alpha}{\Delta(\Delta-\alpha)}+\frac{\alpha}{\Sigma(\Sigma+\alpha)}\right)
\end{equation}

\begin{figure*}[!htb]
\centering

\begin{subfigure}[t]{0.45\textwidth}
    \centering
    \includegraphics[width=\textwidth]{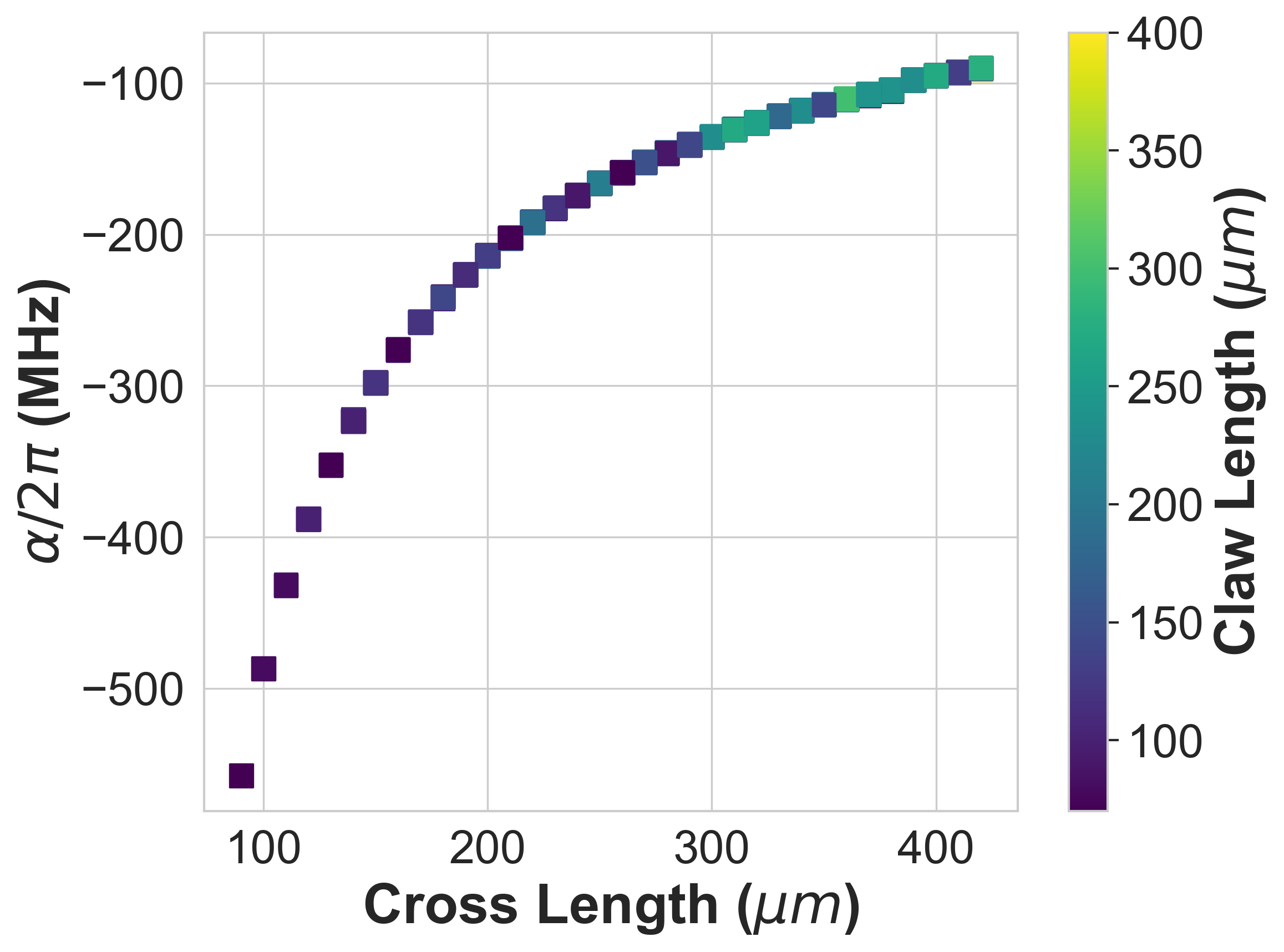}
    \caption{}
    \label{fig:panel1}
\end{subfigure}
\hfill
\begin{subfigure}[t]{0.45\textwidth}
    \centering
    \includegraphics[width=\textwidth]{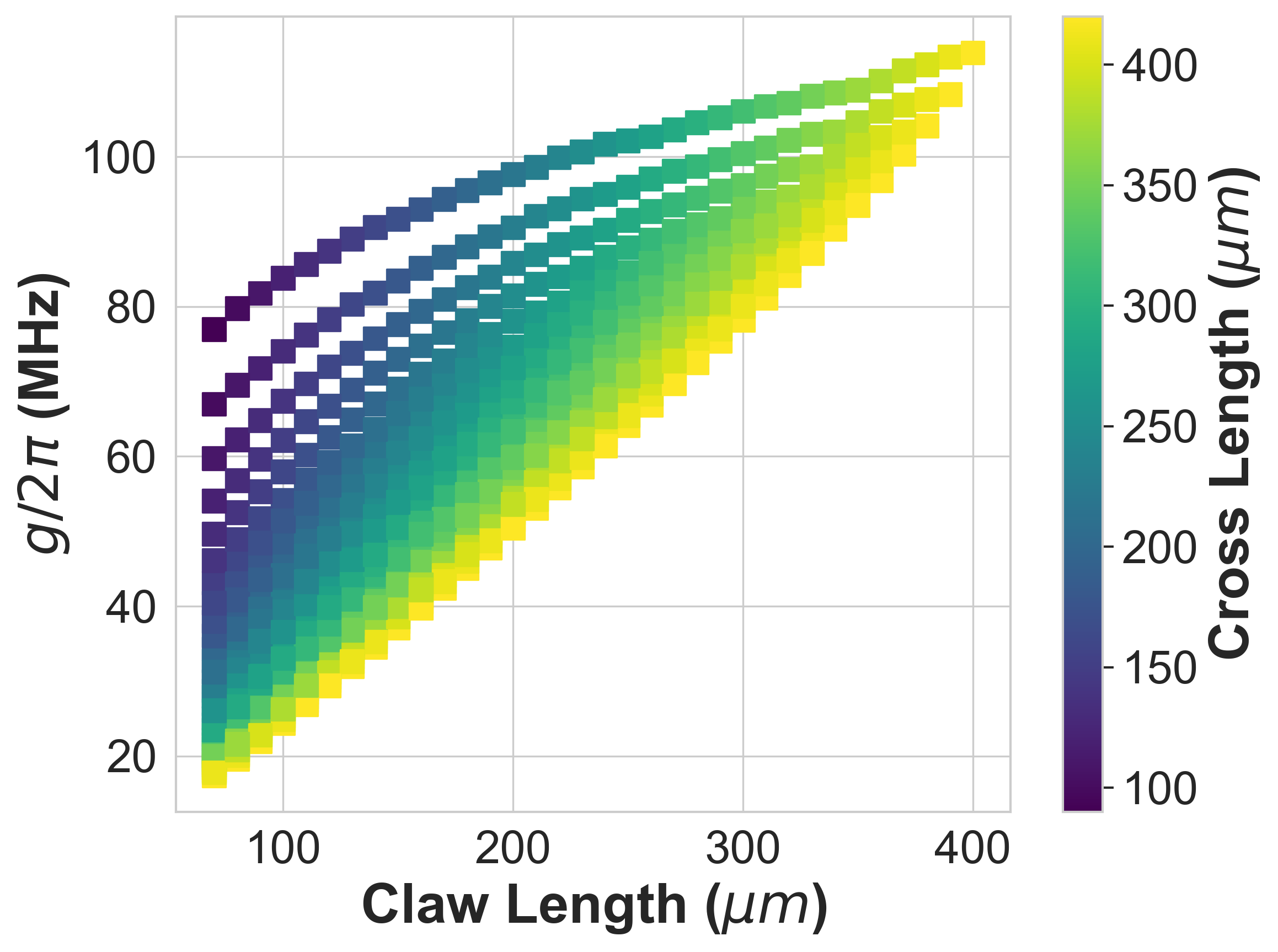}
    \caption{}
    \label{fig:panel2}
\end{subfigure}

\begin{subfigure}[t]{0.45\textwidth}
    \centering
    \includegraphics[width=\textwidth]{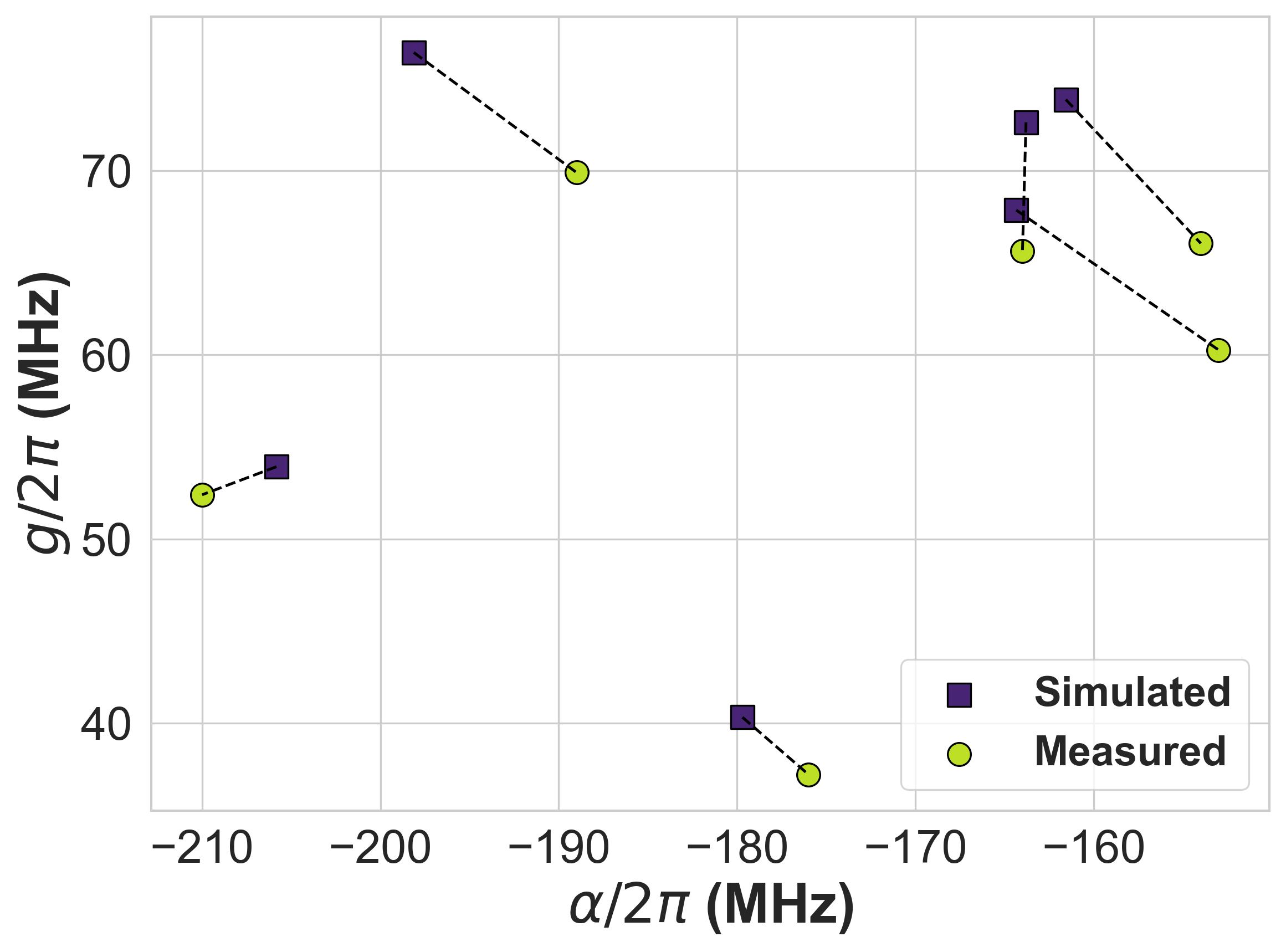}
    \caption{}
    \label{fig:panel3}
\end{subfigure}
\hfill
\begin{subfigure}[t]{0.45\textwidth}
    \centering
    \includegraphics[width=\textwidth]{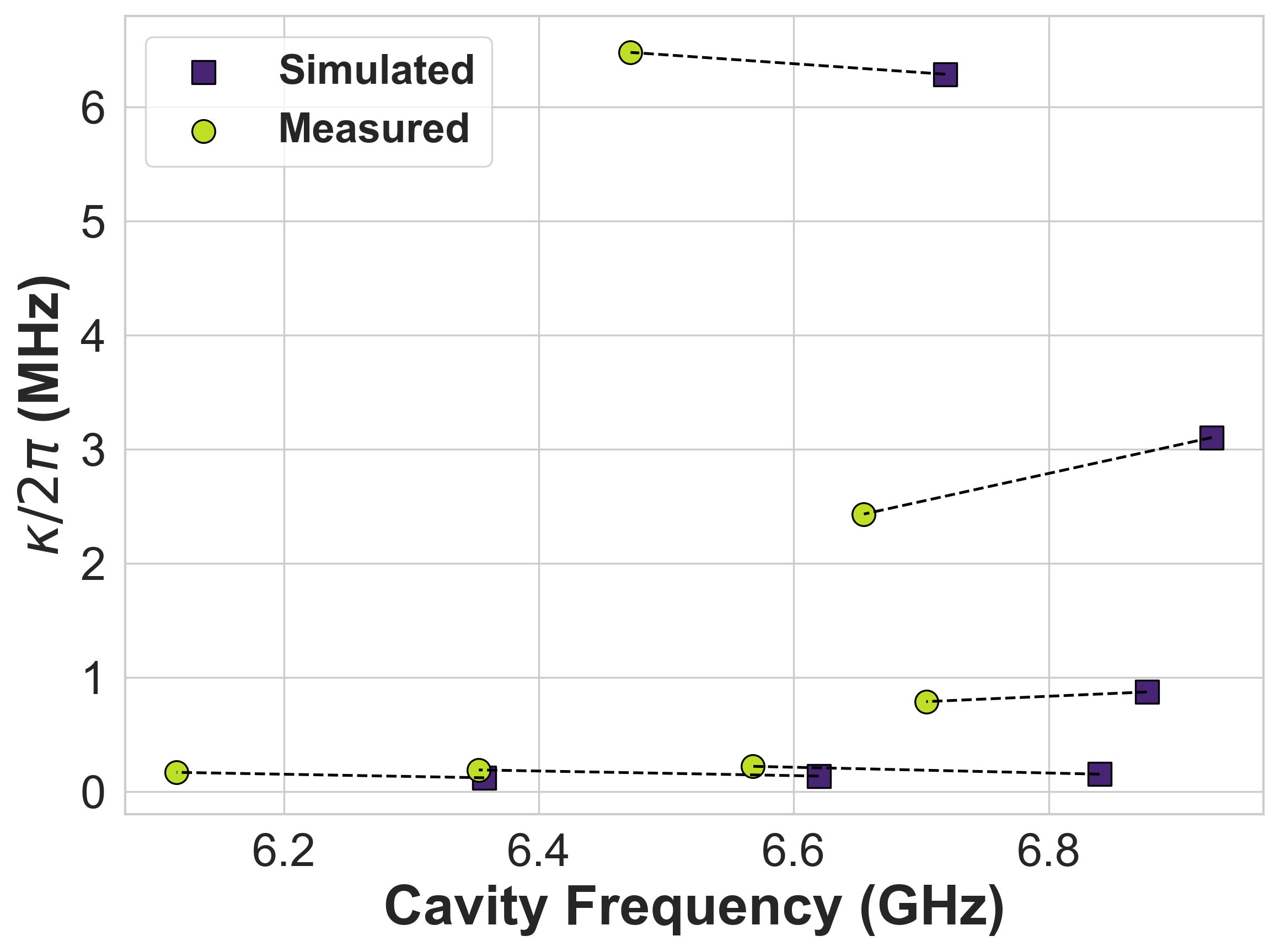} 
    \caption{}
    \label{fig:panel4}
\end{subfigure}

\caption{\justifying
(a) Extracted anharmonicity $\alpha$ as a function of capacitor cross length for simulated xmon devices with various coupling claw lengths. (b) Extracted $g$ for these same xmon devices as a function of coupling claw lengths for a chosen resonator frequency and type. (c) Comparison of experimental measurements and simulated values for the $\alpha$ and $g$ of xmon devices from Fig.~\ref{fig:WM1designPars}. (d) Comparison of experimental measurements and simulated values for resonator frequency $f_r$ and resonator linewidth $\kappa$.
}
\label{fig:HvsDesPars}
\end{figure*}

In this definition $\chi$ is the full resonance shift between qubit ground and excited states (sometimes defined instead as $2\chi$). Again we have included terms typically ignored in the RWA, which give the second term above. This highlights the importance of \emph{not} taking the RWA: if the RWA is taken in the expressions for $\chi_L$ and $\chi$, the $\chi$ predicted from punchout and spectroscopy can be as much as a factor of 2 smaller than the true value! The expressions for $\chi_L$ and $\chi$ are derived from second-order perturbation theory, and are thus subject to corrections of order $g/\Delta$, which is often comparable to experimental precision in any case. See Appendix \ref{app:shifts} for derivations.

We iterate our simulations, refining the HFSS parameters\textemdash mesh size, maximum tolerance, and minimum converged iterations\textemdash until the simulated Hamiltonian parameters converge to the measured ones within reasonable tolerance. Crucially, we ensure that the \emph{same} HFSS parameters yield simulations which are accurate for a range of similar but not identical devices. We have found in the past that it is possible to ``overfit'' a simulation, giving an illusion of accuracy that correctly matches one device's parameters but failing on similar devices. By testing the same simulation parameters against several devices, we ensure their robustness. We achieve good convergence with both analysis methods, but employ lumped-element analysis whenever possible. We make this choice so that later we can build up our database combinatorially. We use eigenmodal analysis only to find the frequencies and coupled quality factor (i.e., $\kappa$) for resonators. We then separately simulate xmons and coupling claws using Q3D to extract capacitances, and use lumped-element formulae and numeric calculations from scqubits to find the Hamiltonian parameters. More details of this procedure are given below in Sec.~\ref{sec:genDB}. Using this procedure, for the devices tested we achieve RMS errors of  $4.1 \%, 16.9 \%, 10.4 \%,$ and $3.8 \%$ for anharmonicity $\alpha$, resonator linewidth $\kappa$, coupling rate $g$, and resonator frequency $f_r$, respectively (see Table \ref{tab:wm1_rms} and Figure \ref{fig:HvsDesPars}).

\begin{table}[]
    \centering
    \begin{tabular}{|c|c|}
        \hline
        Parameter & RMS Error (\%) \\
        \hline
         $\alpha$ & 4.1 \\
        \hline
        $\kappa$ & 16.9 \\
        \hline
        $g$ & 10.4 \\
        \hline
        $f_r$ & 3.8 \\
        \hline
    \end{tabular}
    \caption{RMS errors of the simulated vs. experimentally measured values for various Hamiltonian parameters, for the 6 qubit-resonator combinations shown in Figure 1.}
    \label{tab:wm1_rms}
\end{table}

The somewhat larger error in $\kappa$ is mostly driven by devices with large linewidth ($\kappa \gtrsim 2$ MHz) and is likely the result of impedance variations in the experimental setup which can distort the resonator lineshape and alter the linewidth. We note that the simulations tend to overestimate anharmonicity $|\alpha|$ and consistently overestimate coupling strength $g$. We believe the former effect stems from the assumption of ideal Josephson nonlinearity, i.e., assuming the Josephson energy contribution is exactly $E_J \cos \phi$. This assumption may not be fully valid\textemdash recent work has shown that linear inductance in the transmon circuit and non-ideal behavior in the junction itself can significantly reduce the anharmonicity for a real device with a given $E_C$ \cite{willschObservationJosephsonHarmonics2023}. The overestimation of $g$ is likely in part due to etch bias in the fabrication process, which results in metal features which are narrower than designed. We expect this to contribute a roughly $3 \%$ overestimate of $g$. The overestimate may also be due in part to the omission of air-bridge crossovers from our simulations; these crossovers increase the total resonator capacitance and slightly decrease the coupling capacitance, again reducing $g$. We omit them from the database as their design details are propriety and confidential to the MIT Lincoln Lab foundry that fabricated the device. 

\subsection{Generating the database}\label{sec:genDB}
Once we can be confident in the accuracy of our simulations, we begin simulating devices similar to the experimentally-validated designs, with a range of variations on design geometry parameters. For instance, we simulate devices based on the xmons shown in Fig.~\ref{fig:WM1designPars} with a range of xmon cross length, coupling claw length and width,  resonator CPW line length, and resonator coupling element dimensions. In addition to this device, we simulate variations on the devices shown in \cite{mauryaOndemandDrivenDissipation2024} (with 2 qubit-cavity pairs) and \cite{gaikwadEntanglementAssistedProbe2024} (with 3 qubit-cavity pairs). As we choose a very fine mesh and a low solution tolerance in order to ensure good simulation accurancy, these simulations could be quite time consuming. However, we take advantage of the fact that lumped-element analysis allows us to treat device components as modular lumped elements which can then be combined together, as was done in the simulations of experimentally-measured devices. In this way we can simulate small components and then use analytical formulae to calculate Hamiltonian parameters when they are combined. For instance, we simulate relatively-compact xmon crosses and coupling claws, which only takes a few minutes per simulation. We then can combine these simulated elements with separately-simulated readout cavities, and calculate $g$ from the analytical formula (Eq. \ref{eq:g}). This allows us to grow the database combinatorially, greatly reducing both the number of simulations required and the time required for each simulation. We have confirmed the accuracy of this procedure by testing with validated designs and find that we match experimental parameters well, as discussed earlier. Note that these combinatorial values depend on $E_J$, which is calculated from the user-specified target values of $\omega_q$ and $\alpha$; the database may thus be seen as a set of combinatorial circuits, whose Hamiltonian properties can be calculated once $E_J$ is specified. At present, the database has simulations of 1934 cross-claw structures.

Unfortunately, simulations of distributed-element resonators are still rather lengthy (taking up to an hour or more when performed on a workstation equipped with an AMD Ryzen Threadripper PRO 3955WX 16-core 3.90 GHz processor and 128 GB of RAM), and parameters such as resonator linewidth are not easily calculated from modular component properties for all geometries of the element coupling the resonator to its feedline. For instance, in the ``parallel CPW'' coupling geometry used in 3 of the resonators shown in Fig.~\ref{fig:WM1designPars}, the coupling element is a significant fraction of the length of the resonator and so is difficult to treat modularly \cite{besedinQualityFactorTransmission2018}. For these geometries we therefore run eigenmodal simulations of the full resonator, including the element that couples it to the feedline and a portion of the feedline itself. The feedline is terminated with lumped resistive elements (usually 50 $\Omega$) to simulate the environmental impedance. We also include various combinations of qubit coupling claws, as these affect both resonator frequency and linewidth. From these simulations we directly extract both resonant frequency and external quality factor (i.e., linewidth). These simulations are far more time-consuming and cannot be easily combined combinatorially. We therefore run them in parallel on workstations and on a local high-performance computing cluster. At present, the database has simulations of 693 different quarter-wave resonators with distributed coupling elements. By combining with qubit cross-claw designs, this produces 18,957 devices (where designs are combined so that the claw geometries match).

\begin{figure}
\includegraphics[width=3.3in]{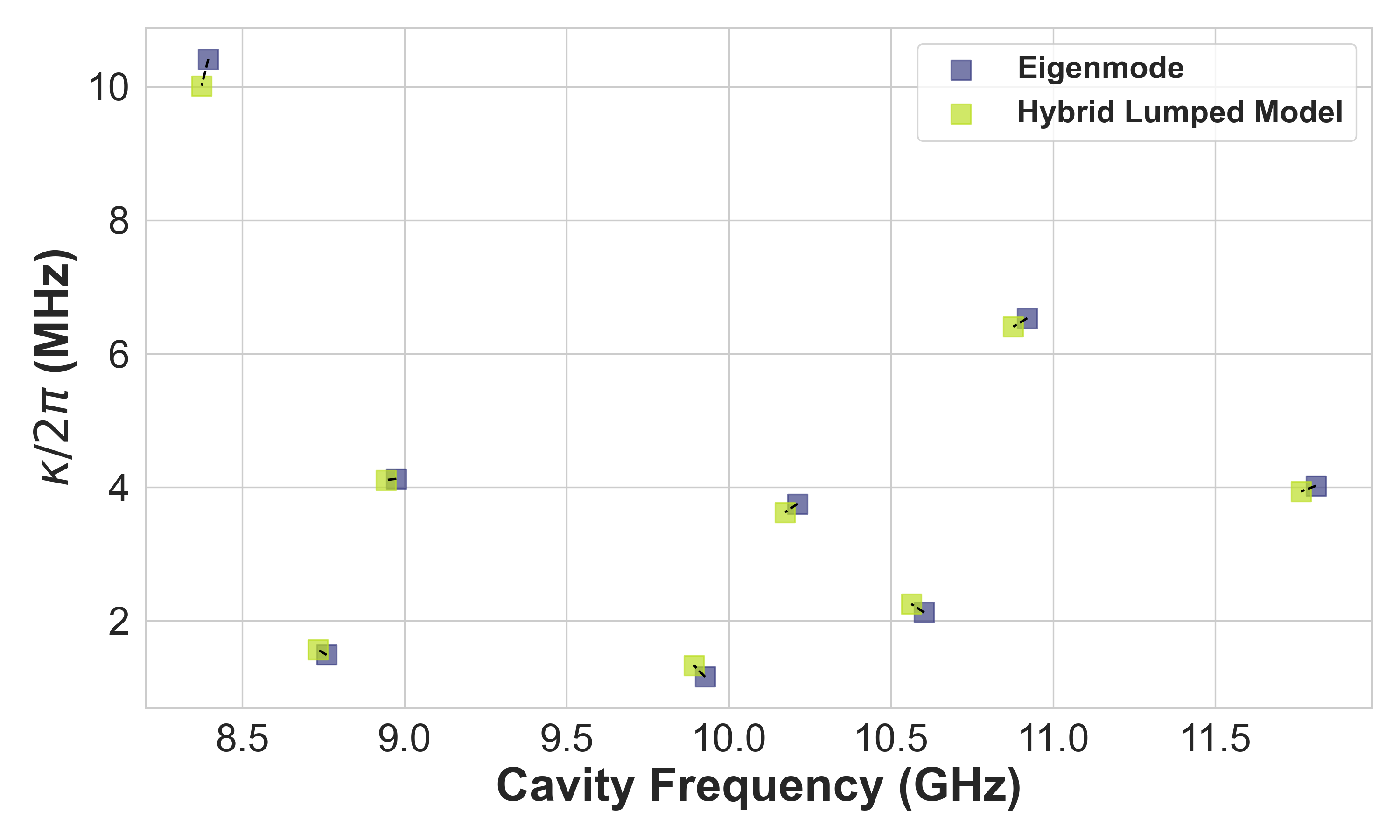}
    \caption{\label{fig:Q3DvsEigenmode}
    Comparison of resonator frequency and linewidth $\kappa$ values obtained via eigenmodal analysis (blue) and via combinatorial lumped-element models (yellow). The two are nearly indistinguishable, showing the validity of the lumped-element approach.
    }
\end{figure}

For resonators with lumped-element capacitive couplers, we first perform eigenmodal simulations of various combinations of CPW length and qubit-coupling claw dimesions with no feedline coupling element to extract the uncoupled resonant frequency $\omega^\prime$. From this we extract the effective resonator capacitance $C_{r}$. We separately perform simulations of the small coupling capacitor and extract its coupling to ground $C_{cg}$ and to the feedline $C_{rf}$. We then calculate the coupled resonant frequency $\omega_r = \sqrt{\frac{C_r}{C_r+C_{rf}+C_{cg}}} \omega^\prime$ and the linewidth $\kappa = \frac{1}{2} Z_0 \omega_r^2 \frac{C_{rf}^2}{C_r+C_{rf}+C_{cg}}$ for all combinations of coupler and CPW/claw, where $Z_0$ is the external environmental impedance (typically 50 or 25 $\Omega$ depending on the geometry). In all cases we automatically extract circuit-level Hamiltonian parameters from the simulation results and store a database entry with the geometry parameters, simulation parameters, and extracted Hamiltonian parameters. We have verified that this produces results for resonator frequency and linewidth that are equivalent to those produced from a full-circuit eigenmodal analysis to within a scatter of a few $\%$---see Figure \ref{fig:Q3DvsEigenmode}. We expect this approach to break down when resonator quality factor $Q$ drops below $\sim 1000$, as the large coupling capacitor would cause distortions of the resonant mode that a lumped model would not capture. We note that this hybrid lumped-distributed approach is qualitatively similar to that taken in \cite{minevCircuitQuantumElectrodynamics2021}, although the details of the models differ. At present, the database has simulations of 406 half-wave resonators with 430 coupling capacitors, yielding 174,580 possible resonator designs; combining with qubit structures yields 4,756,660 device designs.

Fig.~\ref{fig:HvsDesPars} shows how Hamiltonian parameters vary with geometry for the xmon devices discussed above. We achieve good agreement with experimental results and we both interpolate between and extrapolate around the experimental device designs. The largest deviation between simulation and experiment comes in the simulation of the linewidth of devices that are strongly coupled to the feedline, i.e., devices with broad linewidths. We attribute this deviation to impedance variations in the experimental environment that distort the resonance lineshape and modify the linewidth.

\section{User Interface}
\begin{figure}[htbp]
\centering
\resizebox{\columnwidth}{!}{

\begin{tikzpicture}[
    node distance=2cm and 1.5cm,
    auto,
    thick,
    main node/.style={
        draw,
        rectangle,
        rounded corners,
        align=center,
        minimum height=140pt,
        font=\bfseries\fontsize{48pt}{40pt}\selectfont,
        inner sep=8pt, 
        fill=white,
        line width=3pt
    },
    arrow/.style={
        -Latex,
        line width=4pt
    },
    every text node part/.style={align=center} 
]

\node[main node] (connect) {Connect to the \textbf{SQuADDS Database}};
\node[main node, below=of connect] (select) {Select Circuit QED Elements};
\node[main node, below=of select] (input) {Input Target \\ Hamiltonian Parameters};
\node[main node, below left=of input] (receive) {Receive Closest Presimulated Design};
\node[main node, below right=of input] (receiveInter) {Receive Interpolated Design \\ and Simulation Hyperparameters};
\node[main node, below=of receive] (continue) {Continue with Presimulated Design \\ in Qiskit Metal};
\node[main node, below=of receiveInter] (simulate) {Simulate the Interpolated Design};
\node[main node, below=of simulate, yshift=-1.5cm] (contribute) {Contribute the results to the \\ SQuADDS Database};

\draw[arrow] (connect) -- (select);
\draw[arrow] (select) -- (input);
\draw[arrow] (input) -| (receive);
\draw[arrow] (input) -| (receiveInter);
\draw[arrow] (receive) -- (continue);
\draw[arrow] (receiveInter) -- (simulate);
\draw[arrow] (simulate) -- (contribute);

\end{tikzpicture}
} 

\caption{\label{fig:FlowChart}\small
Process workflow of the SQuADDS database from a user perspective. The user selects a device type and target parameters, and optionally defines a cost function. SQuADDS searches the database for the closest pre-simulated design and experimentally validated design, and outputs these designs (and specified $E_J$ values) and code to generate and simulate them. It also interpolates a best-guess design and gives the user code to generate and simulate it, along with estimated Hamiltonian parameters.}   
\end{figure}
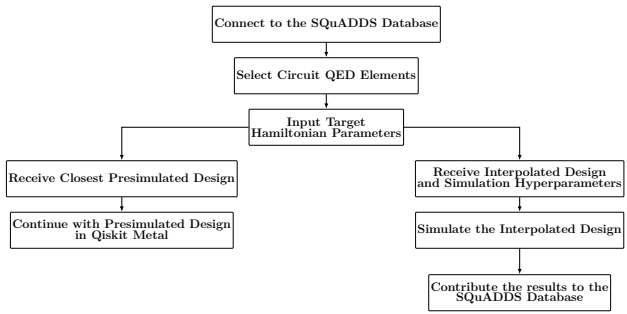

The workflow from a user perspective is diagrammed in Fig.~\ref{fig:FlowChart}. First, the user chooses a set of basic device characteristics, selecting which components are coupled to each other, whether resonators should be half-wave or quarter-wave, what form the resonator-feedline coupling should take, and what drive lines should be included. They then are then queried for relevant Hamiltonian parameters: qubit frequency and anharmonicity, resonator frequency and linewidth, qubit-resonator coupling rate, and drive-line-limited qubit lifetime. They can optionally choose their own cost function, selecting how to calculate which design is the ``best match'' to their target parameters, or use the default cost function (details below).When SQuADDS receives this input it first uses scQubits to calculate the required $E_J$ values. It then searches the design database for existing designs with closely matching Hamiltonian parameters, and tags the closest and next-closest designs for each parameter, as well as the closest design overall based on the cost function. It then interpolates device geometry parameters to reach a ``best-guess'' design that has not yet been simulated (details below). The user is presented with this best-guess design and its estimated Hamiltonian parameters, including both a .gds file and code to programmatically generate the layout. This code also includes programmatic launching and running of a finite-element ANSYS HFSS simulation to extract the Hamiltonian parameters of the design. The simulation parameters (such as mesh size and maximum deviation tolerance) are matched to those that had reliably reproduced the experimental results of the closest validated design. The user is also presented with the pre-simulated design that is closest to their specified parameters, along with the corresponding closest experimentally validated design. An example user input and output is provided as supplementary material.

The choice of cost function is key to finding a suitable device. Typically a simple Euclidean square distance measure is suitable, which computes the cost of simulated Hamiltonian parameters $\{p_i\}$ versus target parameters $\{P_i\}$:
\begin{equation*}
    F(\{P_i\},\{p_i\}) = \sum_i w_i\frac{(P_i - p_i)^2}{P_i^2}
\end{equation*}
Here $w_i$ are weights which default to 1 but may be user-defined. For instance, a user may require tight precision of a few percent on the readout resonator frequency $\omega_r$, while tolerating $10-20\%$ deviation from the target qubit frequency $\omega_q$, if they have many readout resonators sharing the same feedline. In this case the user should upweight the resonator frequency error and downweight the qubit frequency error in the cost function. The user can also optionally choose to include functions of the parameters, such as $\Delta$ and $\chi$, in the cost function, and can ignore error in any parameter or function by setting its weight to 0. They can also choose different distance metrics, and if desired can write completely custom cost functions.

\section{Design interpolation}
Our goal is to provide a design database extensive enough that there is always a well-simulated design with Hamiltonian parameters close to a user's goal. However, even a bounded parameter space for simple single-qubit devices is many-dimensional (resonator and qubit frequency, qubit anharmonicity, resonator linewidth, qubit-resonator coupling, etc.) and so covering every combination of parameters with fine resolution is computationally challenging. Instead, we try to cover as wide of a range of design geometry parameters as possible with moderate parameter point spacing, then interpolate between them based on user input. This presents its own challenge, as each design geometry parameter typically influences multiple Hamiltonian parameters. Ordinarily this would be an ideal use case for machine learning, and indeed we have had some success with such approaches in parameter regimes where we have good coverage. However, for the time being our coverage of resonator parameter space is relatively sparse, and so machine learning algorithms tend to overfit and produce inaccurate results. We expect this issue to diminsh as we continue to add designs to the database, and we plan to switch the interpolation logic over to a machine learning model soon.

In the meantime, we can use our knowledge of the physics of the system to provide a physically-motivated interpolation procedure: 

\begin{figure*}[!htb]
\centering
\begin{subfigure}[t]{0.45\textwidth}
    \centering
    \includegraphics[width=\textwidth]{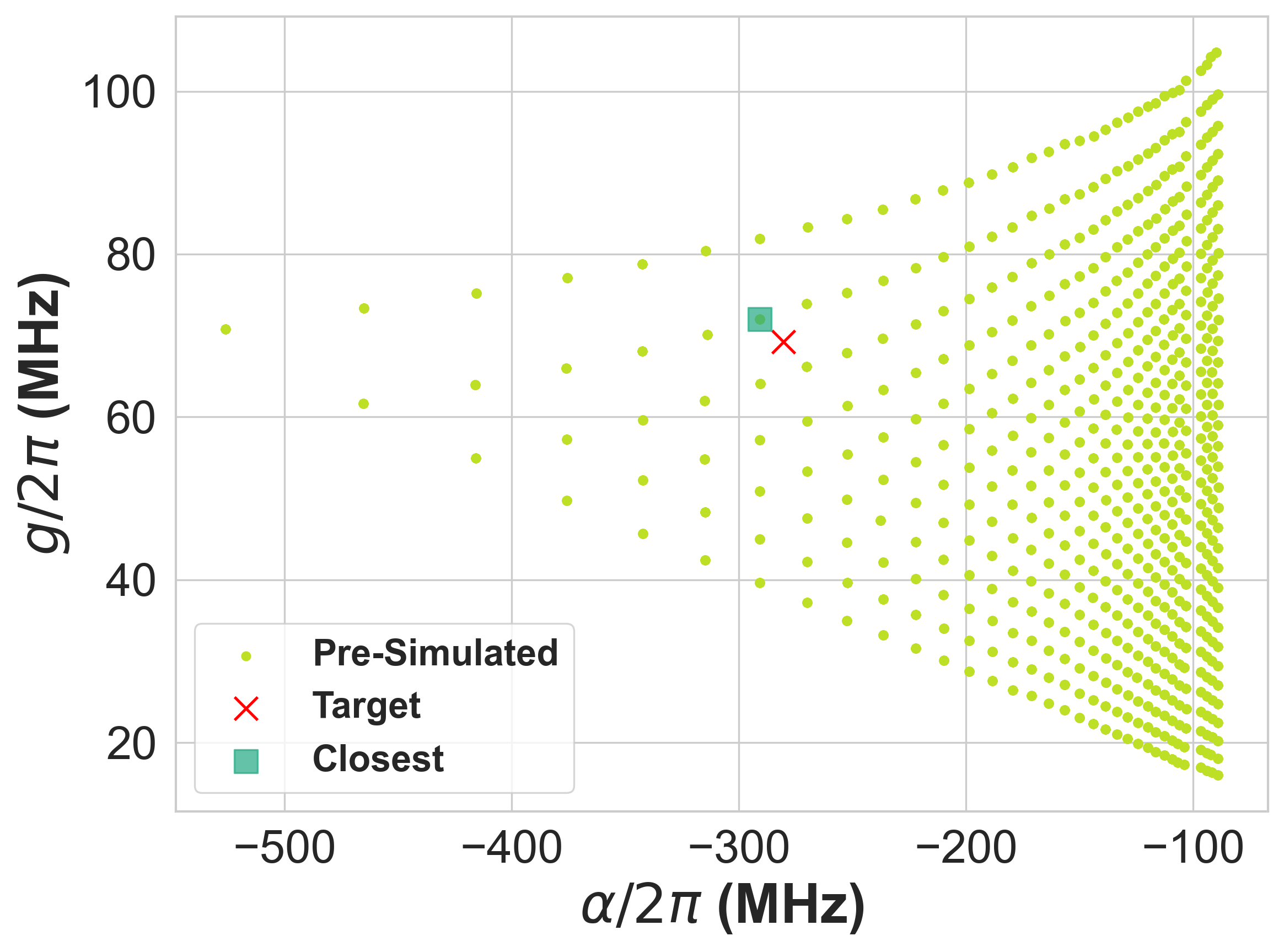}
    \caption{}
    \label{fig:panel1A}
\end{subfigure}
\hfill
\begin{subfigure}[t]{0.45\textwidth}
    \centering
    \includegraphics[width=\textwidth]{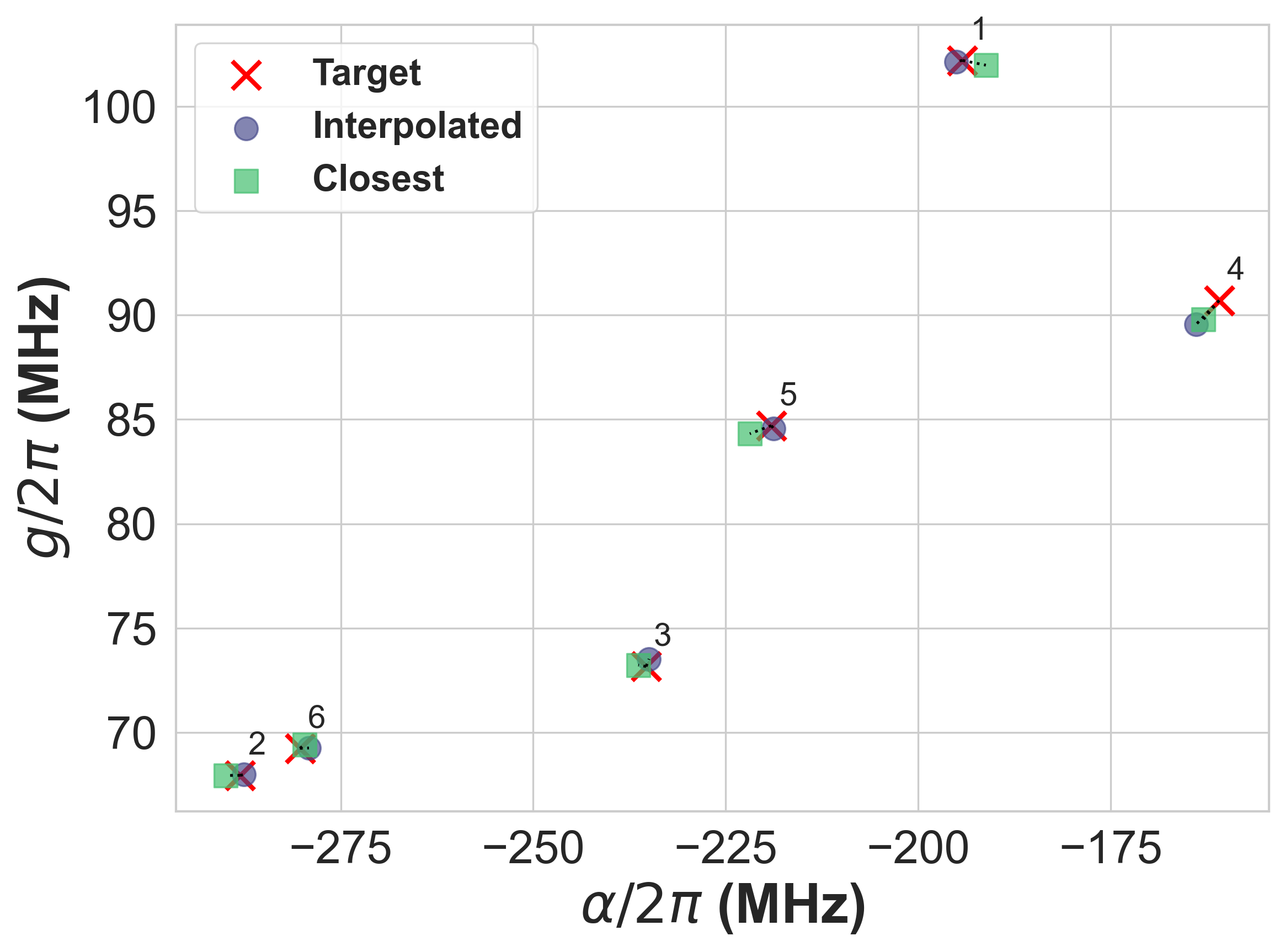}
        \caption{}

    \label{fig:panel2A}
\end{subfigure}

\begin{subfigure}[t]{0.45\textwidth}
    \centering
    \includegraphics[width=\textwidth]{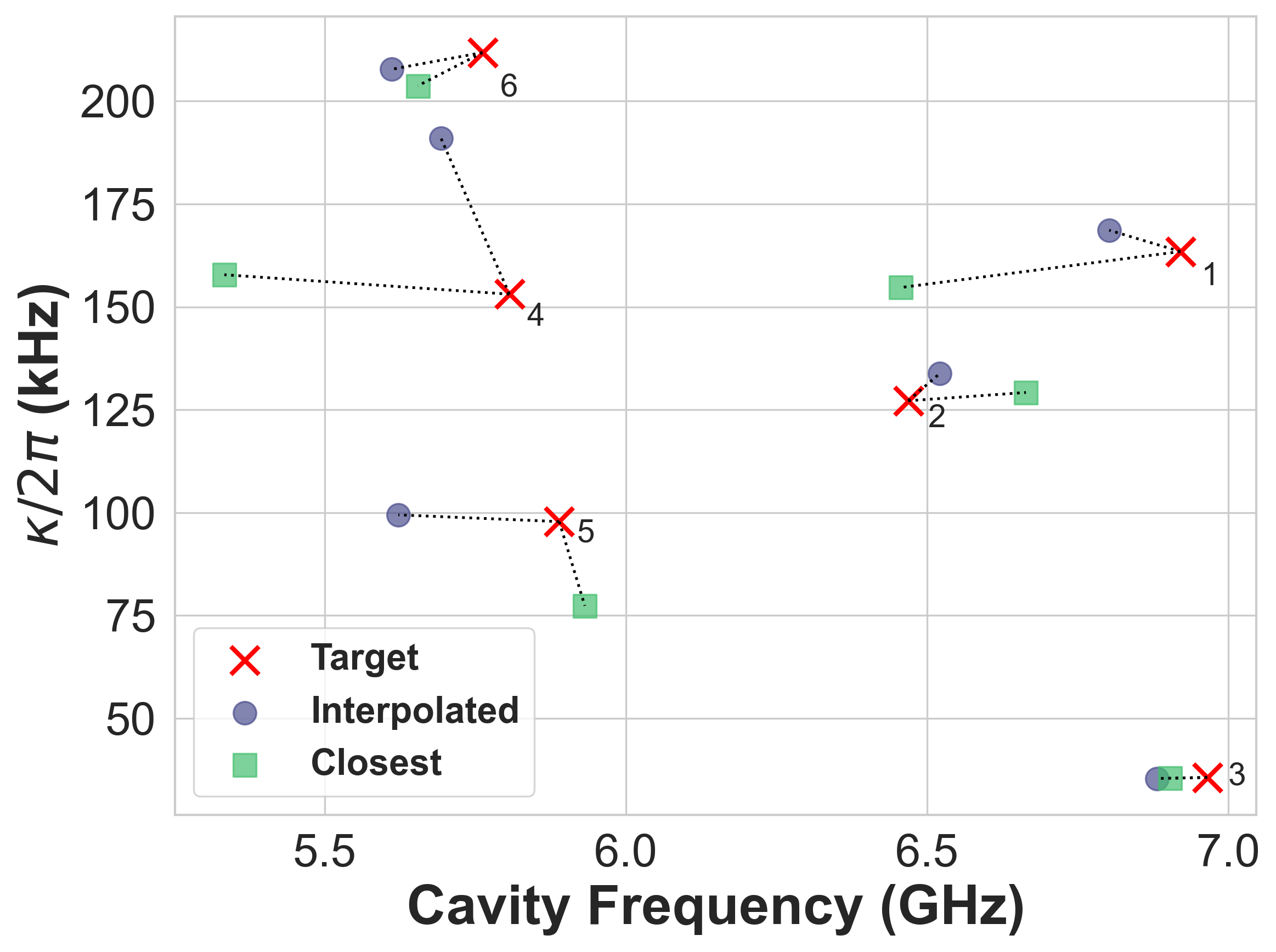}
        \caption{}

    \label{fig:panel3A}
\end{subfigure}
\hfill
\begin{subfigure}[t]{0.5\textwidth}
    \centering
    \vspace{-119pt} 
    \footnotesize
    \setlength{\tabcolsep}{2pt} 
    \renewcommand{\arraystretch}{1.2} 
   \begin{tabular}{c|c|c|c|c|c|c}
    \hline
    Design & $\Delta g (\%)$ & $\Delta \alpha (\%)$ & $\Delta \kappa (\%)$ & $\Delta f_r (\%)$ & $\Delta f_q (\%)$ & $\max_{\{\xi_i\}}(\xi,\Delta \xi)$ \\
    \hline
    1 & 0.05 & 0.42 & 3.17 & 1.71 & 0.38 & ($f_r$, 6.70\%) \\
    2 & 0.05 & 0.16 & 5.24 & 0.81 & 0.12 & ($f_r$, 3.02\%) \\
    3 & 0.49 & 0.16 & 0.70 & 1.20 & 0.12 & ($f_r$, 0.89\%) \\
    \textcolor{red}{4} & \textcolor{red}{1.27} & \textcolor{red}{1.87} & \textcolor{red}{24.78} & \textcolor{red}{1.96} & \textcolor{red}{0.04} & \textcolor{red}{($f_r$, 8.15\%)} \\
    5 & 0.16 & 0.12 & 1.67 & 4.54 & 0.10 & ($\kappa$, 20.91\%) \\
    6 & 0.02 & 0.42 & 1.92 & 2.62 & 0.23 & ($\kappa$, 3.82\%) \\
    \hline
    \end{tabular}
    \vspace{10pt} 
    \caption{}
    \label{fig:panel4A}
\end{subfigure}

\caption{\justifying \label{fig:interp}
    (a) Qubit anharmonicity $\alpha$ and qubit-resonator coupling strength $g$ for various pre-simulated designs with a user-specified target point (red x) and closest pre-simulated design (green square) highlighted. (b) Qubit-resonator coupling strength ($g$) and qubit anharmonicity ($\alpha$) with 6 target points (red crosses) compared to the best pre-simulated points (green squares) and interpolated designs (purple circles). The interpolation improves accuracy in all but one case (rightmost point), where the target lies outside the convex bounds of the pre-simulated space. (c) resonator linewidth ($\kappa$) and resonant frequency ($f_r$), again comparing the best pre-simulated designs and the interpolated designs against the user-defined targets. Again interpolation almost always improves accuracy, although overall accuracy suffers when the target is outside the pre-simulated convex bounds. (d) Table containing the RMS percentage differences from the target values for the interpolated points in the \(g\), \(\alpha\), \(\kappa\), and \(f_r\) space, demonstrating the overall precision of the interpolation process from the SQuADDS database. In general we see excellent accuracy of these designs across all Hamiltonian parameters, except when the target falls outside the convex bounds of the pre-simulated parameter space (Design No. 4). The last column indicates the largest deviation of a target parameter from the closest pre-simulated design.}
\label{fig:accuracy}
\end{figure*}

\begin{enumerate}
    \item From the user-input qubit type and parameters (frequency, anharmonicity, coupling strength, resonator frequency and type), numerically calculate the required qubit capacitance and coupling capacitance, as well as $E_J$. If $E_J/E_C < 30$, warn the user. 
    \item Search the database for the design combination of qubit capacitance and coupling capacitance that gives the best match to the anharmonicity $\alpha_\mathrm{target}$ and coupling strength $g_\mathrm{target}$ given the specified target resonator frequency and qubit frequency, according to the specified cost function. Flag this design.
    \item Scale the qubit capacitor and the coupling capacitor area linearly by the ratio $\alpha_\mathrm{simulated}/\alpha_\mathrm{target}$. Then scale the coupling capacitor area only by the ratio $g_\mathrm{target}/g_\mathrm{simulated}$. 
    \item From the user-input resonator type (quarter-wave or half-wave) and target parameters (frequency and linewidth), select the pre-simulated resonator with the closest combination of values according to the specified cost function. Include only designs with qubit coupling claw capacitances that differ from the value above by less than 30\%. Flag this design.
    \item Take this pre-simulated resonator design and replace its qubit coupling claw with the one calculated in step 3. Scale the resonator length linearly to reach the desired frequency, and scale the square of the feedline coupling element characteristic dimension (e.g., the fingers of an interdigitated capacitor) linearly to reach the desired linewidth. 
    \item If the resonator-feedline coupling element (coupling capacitor or co-linear waveguide section) has changed by more than $1\%$ of the total resonator capacitance (or length), re-calculate the resonant frequency $f_r$ and rescale the resonator length as necessary. For  Likewise if the qubit coupling claw capacitance $C_c$ has changed by more than 1 \% of $C_r$, rescale the resonator length to hit the target frequency.
    \item Return to the user the interpolated design, code to generate and simulate the design, and the designs and code for the closest pre-simulated designs.
\end{enumerate}
A pseudocode version of this algorithm is included in Appendix \ref{app:interp}.

This procedure is of course not fully accurate. Most of the scalings are not truly linear, and there are interactions neglected. For example, changing the qubit-resonator coupling element changes the total qubit capacitance and thus the anharmonicity, which we do not compensate for, under the assumption that the coupling capacitance is small compared to the total capacitance. Given that our coverage of the parameter space with pre-simulated designs is reasonably dense, and will become more so as more designs are added to the database, we expect all scaling factors to be relatively close to 1. We therefore do not expect these approximations to cause major inaccuracies.

Tests of this procedure indicate its reliability within the convex bounds of the pre-simulated parameters space. See Fig.~\ref{fig:accuracy}. We have selected various points in the parameter space and then simulated the interpolated designs; on average they have rms deviation of $0.56\%, 0.81\%, 2.46\%, 10.47\%, 0.20\%$ from the target $g, \alpha, f_r, \kappa,$ and $f_q$ respectively. Almost all of the $\kappa$ error comes from one point (Design No. 4 of table in Fig.~\ref{fig:interp}(d)). where the target was far outside the convex bounds of the presimulated space, and so interpolation is less reliable; we expect this issue to diminish as we continue adding designs.  Of course, a user with simulation capability can always simulate the design to confirm its parameters; we generate the code necessary to ensure an accurate simulation as part of the output.

We note that this interpolation algorithm is quite specific to our particular design. To perform interpolation more generally, we are working to implement machine learning algorithms that are trained on our pre-simulated data sets. Preliminary results are promising, but more database coverage is required before the new interpolation algorithms reach the accuracy of the current physics-based algorithm. Adding this database coverage and training general interpolation models are planned for the coming months.

\section{Future development}\label{sec:future}

The SQuADDS project currently has some limitations. The database is limited in extent, with only xmon-style transmons coupled to quarter- and half-wave CPW resonators, all with somewhat similar dimensions. We plan to continue adding validated designs to the database every time a new device is measured in our group. More importantly, the open-source nature of the project means that users from other groups can contribute their experimentally-measured designs, thus providing a much broader dataset. These experimentally-validated designs then provide "fixed points" in design geometry parameter space, about which we and other contributors can extrapolate geometries and simulate the resulting designs. All contributed designs will be subject to an approval process including review of raw data, ensuring that SQuADDS remains an accurate resource. Additionally, the database structure of SQuADDS makes it straighforward to include more information about validated devices, such as fabcrication recipes, device images, materials properties, paper references, etc.

Users will also have the opportunity to contribute to the SQuADDS codebase through pull requests to the project GitHub repository \cite{GITHUB}. A wishlist of features to add is maintained on the repository. Features we plan to add in the near term include functionality to calculate fluxonium qubit parameters, multi-qubit and Purcell-filtered designs, and functionality to compute Hamiltonians of arbitrary circuits based on the physical layout (by interfacing with packages such as SQcircuits). We furthermore plan to change Hamiltonian parameter calculation from using analytical formulas (which are typically only accurate in the limit of weak coupling) to using scqubits and/or SQcircuits to numerically calculate all parameters.

Currently the simulation workflow is based on the proprietary ANSYS HFSS solver, which can be quite expensive, especially for high-performance computing licenses. This difficulty is significant as not all groups who wish to contribute may have access to HFSS, and our own group is mostly limited to running simulations on in-lab workstations rather than a computing cluster. In the next update to SQuADDS, we plan to integrate support for the AWS PALACE open-source solver, which resolves both the accessibility and cost issues \cite{PALACE}.

In an effort to make SQuADDS a more versatile tool for both quantum hardware developers and machine learning (ML) researchers, we utilize the HuggingFace platform to host our database \cite{SQuADDSDB}. This approach facilitates the study of our results through ML models, akin to those documented in references \cite{Nugraha2023MLPredictiveModel, Zhang2018MultivaluedNN, Feng2022ANNMicrowaveCAD}. Having our database accessible to train future models, we hope to aid in the development of Electronic Design Automation (EDA) tools that can catalyze innovation in the superconducting quantum hardware, industry mirroring the pivotal impacts seen in the semiconductor field \cite{Huang2021MachineLearningEDA}. 

In conclusion, we have created SQuADDS, a resource for the superconducting qubit community that enables researchers to quickly and easily generate device designs from desired Hamiltonian parameters. The project is underpinned by a database of pre-simulated designs, which are variations on simulations that have been empirically matched to experimental results. The project's open-source ethos invites active community engagement to expand the database, introduce new functionalities such as additional qubit types, novel geometric design elements, and calculations of qubit-qubit and qubit-resonator couplings, as well as to enhance the user interface. Our project significantly lowers the barrier to entry for new research groups and reduces the startup costs for existing groups to experiment with innovative device designs.

\begin{acknowledgments}
We gratefully acknowledge Zlatko Minev, Jens Koch, Cyrus Hirjibehedin, and the IBM Qiskit Metal team for useful discussions, and the Qiskit user community for help with software issues. Devices were fabricated and provided by the Superconducting Qubits at Lincoln Laboratory (SQUILL) Foundry at MIT Lincoln Laboratory, with funding from the Laboratory for Physical Sciences (LPS) Qubit Collaboratory. This work was supported by NSF grant OMA-1936388, ONR grant N00014-21-1-2688, and RCSA Cottrell grant 27550. 
\end{acknowledgments}


\input{main.bbl}

\appendix
\renewcommand{\thefigure}{S\arabic{figure}}
\renewcommand{\thetable}{S\Roman{table}}

\section{Lamb and Dispersive Shift Derivations}\label{app:shifts}
We begin by treating the uncoupled transmon-resonator Hamiltonian as the unperturbed system:
\begin{equation}
    H^0 = \omega_r (a^\dag a + 1/2) + \omega_q b^\dag b + \frac{\alpha}{2} b^\dag b(b^\dag b-1)
\end{equation}
with eigenstates $\ket{m n}$ and eigenenergies $E^0_{m n}=(m+\frac{1}{2})\omega_r + n\omega_q + \frac{\alpha}{2}(n^2-n)$
We next introduce the perturbation of the transmon-resonator interaction:
\begin{equation}
    H_{int}\approx g(a-a^\dag)(b-b^\dag)
\end{equation}
We next compute the corrections to the energies due to $H_{int}$. There is no first-order correction, as $E^1_{m n} = \bra{m n}H_{int}\ket{m n} = 0$. The second-order expression is:
\begin{align}
    &E^2_{mn} = \\ \notag
    &\sum_{m^\prime n\prime \neq m n} \frac{|\bra{m^\prime n^\prime}g(a b-a^\dag b - a b^\dag + a^\dag b^\dag)\ket{mn}|^2}{E^0_{m n} - E^0_{m^\prime n^\prime}} \notag
\end{align}
The only terms which are nonzero are those with $(m^\prime, n^\prime) = \{(m-1, n-1);(m-1,n+1);(m+1,n-1);(m+1,n+1)\}$. These terms give:
\begin{align}
    &E^2_{mn} =g^2\left( \frac{mn}{\Sigma + (n-1)\alpha}     + \frac{m(n+1)}{\Delta - n\alpha}\right. \\ \notag
    &\left. - \frac{(m+1)n}{\Delta - (n-1)\alpha} - \frac{(m+1)(n+1)}{\Sigma + n\alpha}\right) \notag
\end{align}
where again the first and last term are due to the non-RWA terms that are dropped in many derivations, $\Delta \equiv \omega_r - \omega_q$, and $\Sigma \equiv \omega_r + \omega_q$.

The Lamb shift is given by $\chi_L = E^2_{(m+1)0} - E^2_{m0}$---the change in resonator energy level splitting due to the presence of the qubit in its ground state. This gives
\begin{align} \label{eq:lamb}
    \chi_L = \frac{g^2}{\Delta} - \frac{g^2}{\Sigma}
\end{align}
as stated in the main text.

The dispersive shift is given by $\chi = (E^2_{(m+1)1} - E^2_{m1})- (E^2_{(m+1)0} - E^2_{m0})$---the change in resonator energy level splitting when the qubit changes from the ground state to the excited state. Plugging in yields
\begin{align}
    \chi =& 2g^2\left(\frac{1}{\Sigma}+\frac{1}{\Delta-\alpha}-\frac{1}{\Delta}-\frac{1}{\Sigma+\alpha}\right) \\ \notag
    =& 2g^2 \left( \frac{\alpha}{\Delta(\Delta-\alpha)} + \frac{\alpha}{\Sigma+\alpha}\right)
\end{align}
as stated in the main text. The terms involving $\Sigma$ come from the Hamiltonian terms that are thrown away in the RWA, but they are often not negligible.

\section{Interpolation Algorithm}\label{app:interp}
Below we give a pseudocode version of the interpolation algorithm.


\begin{algorithm}[tb]

\caption{Interpolation for Qubit-resonator Design}
\vspace{0.5cm}
\KwIn{Qubit type and parameters: $\omega_q$, $\alpha$, $g$, $\omega_r$, \textit{resonator\_type}}
\KwOut{Interpolated design and simulation code}
\SetAlgoLined
\begin{enumerate}
  \item Calculate $C_q$, $C_c$, and $E_J$ from user input
  \item \textbf{If }{$E_J / E_C < 30$} \textbf{then} Warn the user
  \item Select the qubit-claw design with the best match to $\omega_q$, $\alpha$, $g$ given $\omega_r$
  \item Scale the qubit-claw design by scaling the cross length and claw length:
  \begin{itemize}
    \item $l_{xmon} \rightarrow l_{xmon} \times \left(\frac{\alpha_\text{sim}}{\alpha}\right)$
    \item $l_{claw} \rightarrow l_{claw} \times \left(\frac{\alpha_\text{sim}}{\alpha}\right) \left(\frac{g}{g_\text{sim}}\right)$
  \end{itemize}
  \item Filter the database to include only resonator-claw designs with $C_c$ within 30\% of the chosen qubit-claw design
  \item Select the resonator-claw design that gives the closest match to $\omega_r$, $\kappa$ given the \textit{resonator\_type}
  \item Update the claw length of the selected resonator-claw design with that from step 3 and scale the resonator length linearly:
  \begin{itemize}
    \item $l_{res} \rightarrow l_{res} \times \left(\frac{\omega_r}{\omega_{r,\text{target}}}\right)$
  \end{itemize}
  \item Scale the square of the feedline coupling element characteristic dimension of the resonator-claw design:
  \begin{itemize}
    \item $l_{fline} \rightarrow l_{fline} \times \sqrt{\frac{\kappa}{\kappa_\text{target}}}$
  \end{itemize}
  \item \textbf{If }{$C_{res-fline}$ or $C_c$ changes exceed 1\% of total $C_{res}$} \textbf{then} Recalculate $\omega_r$ and adjust $l_{res}$
  \item Return the interpolated design, code to generate and simulate the design, and the designs and code for the closest pre-simulated designs
\end{enumerate}
\end{algorithm}


\end{document}

%% file: main.bbl
\newcommand{\DoNotMakeWarningsErrors}{}
\DoNotMakeWarningsErrors